\documentclass[12pt]{iopart}
\usepackage{iopams}  
\usepackage{graphicx}
\def\sub#1{_{\mbox{\scriptsize{#1}}}}
\def\ie{\textit{i.e.} }
\def\etal{\textit{et al.} }
\def\eg{\textit{e.g.} }
\def\mc#1{\mathcal{#1}}

\def\re#1{(\ref{#1})}
\def\bkt#1{\left(#1\right)}
\def\erfc{\mbox{erfc}}

\newcommand{\sig}{\sigma}
\newcommand{\dt}{\delta}
\newcommand{\eps}{\epsilon}
\newcommand{\ff}{\phantom{t}}

\newcommand{\mb}{\mathbf}

\newcommand{\be}{\begin{eqnarray}}
\newcommand{\ee}{\end{eqnarray}}
\newcommand{\nn}{\nonumber}
\newcommand{\bit}{\begin{itemize}}
\newcommand{\eit}{\end{itemize}}

\def\mpc{Mpc$^{-1}$ }

\newcommand{\C}{\xi}

\newcommand{\LL}{^\ell\lambda_H}

\newcommand{\M}{m_{\mbox{\scriptsize{Pl}}}}

\def\dc{\delta_{c}}
\def\opbh{\Omega\sub{pbh}}
\def\pbh{primordial black hole }

\def\pbhs{primordial black holes }

\def\rt{\right}
\def\lt{\left}
\def\pr{\mc{P_R}}
\begin{document}

\title[Accuracy of slow-roll formulae: implications for primordial black hole  formation.]{Accuracy of slow-roll formulae for inflationary perturbations: 
implications for primordial black hole formation. } \author{Sirichai
Chongchitnan and George Efstathiou}

\address{Institute of Astronomy. Madingley Road, Cambridge, CB3 OHA. United Kingdom.}
\ead{sc427@cam.ac.uk}
\begin{abstract}

We investigate the accuracy of the slow-roll approximation for
calculating perturbation spectra generated during inflation. The
Hamilton-Jacobi formalism is used to evolve inflationary models with
different histories. Models are identified for which the scalar power
spectra $\pr$ computed using the Stewart-Lyth slow-roll approximation
differ from exact numerical calculations using the Mukhanov
perturbation equation. We then revisit the problem of primordial black
holes generated by inflation. Hybrid-type inflationary models, in
which the inflaton is trapped in the minimum of a potential, can
produce blue power spectra and an observable abundance of primordial
black holes. However, this type of model can now be firmly excluded from 
observational constraints on the scalar spectral index on cosmological
scales. We argue that significant primordial black hole formation in
simple inflation models requires contrived potentials in which there is a period of fast roll  towards the end of inflation. For this type of model, the 
Stewart-Lyth formalism breaks down. Examples of such inflationary
models and numerical computations of their scalar fluctuation spectra
are presented.

\end{abstract} 

\maketitle

\section{Introduction}

The simplest models of inflation predict an almost scale-invariant and Gaussian distribution of primordial density perturbations as a consequence of the inflaton field rolling slowly down a potential of a specific form  \cite{albrecht,linde}. Provided that the rolling is sufficiently slow, a set of `slow-roll' approximations allows the primordial perturbations to be computed easily. 

However, there is no compelling reason why inflation should be of the slow-roll type. Inflationary scenarios that violate the slow-roll conditions have been studied by many authors  \cite{linde2,adams,dimopoulos,starobinskii,matsuda}. Furthermore, observations that constrain the amplitude of primordial perturbations only probe a relatively small range of scales. It is therefore possible that inflation might have been more unusual than that anticipated in the simplest theories, and that there may have been a period of fast roll before inflation ended.

In this paper, we analysed inflationary perturbations numerically for single-field inflationary models with generic evolutionary histories, including cases where slow-roll conditions are badly violated. We then compare the numerical results with those calculated using slow-roll formulae, and identified cases where the two approaches differ. 

We revisit the problem of primordial black holes generated by inflation. We argue that primordial black hole formation in slow-roll inflation models is excluded observationally.  It is, however, possible to form primordial black holes if there was a period of fast roll towards the end of inflation. For this type of model, numerical computation of the fluctuations is essential to calculate the abundance of primordial black holes. In the final section, we give an explicit example of a model which satisfies observational constraints on CMB scales, but produces  primordial black holes in significant abundance.  The model is, of course, contrived, but gives an indication of the level of fine-tuning needed to produce \pbhs in single-field inflation.

\section{Evolution of scalar perturbations}

\subsection{Generating inflationary models}

Given our lack of knowledge of the fundamental physics underlying
inflation, we shall investigate scalar perturbations in a large number
of inflationary models numerically. Our approach is based on the
Hamilton-Jacobi formalism introduced in \cite{hoffman}. Tensor perturbations were
treated using this formalism in our earlier work \cite{me2}.

In this approach, the dynamics of inflaton $\phi$ is determined by the
Hubble parameter $H(\phi)$, which is related to the
potential $V(\phi)$ via the Hamilton-Jacobi equation 
\be
\left(H'(\phi)\right)^2-{12\pi\over{\M^2}}H^2(\phi)=-{32\pi^2\over\M^4}V(\phi)\ff.\label{ham}\ee
The functional form of $H(\phi)$ can be specified by a hierarchy of
`flow' parameters: \be \eps &\equiv& {\M^2\over
4\pi}\left({H'\over H}\right)^2,\ff\qquad\qquad \eta \equiv {\M^2\over
4\pi}\lt({H''\over H}\rt)\ff,\nn\\ \xi&\equiv&\lt({\M^4\over
16\pi^2}\rt)\bkt{ H^\prime H^{\prime\prime\prime}\over
H^2}\ff,\ff\ff\ff\sig \equiv 2\eta -4\epsilon\ff,\label{flowparam}\\
^\ell\lambda_H\ &\equiv& \bigg({\M^2\over
4\pi}\bigg)^\ell{(H')^{\ell-1}\over H^\ell}{d^{\ell+1}H\over
d^{\ell+1}\phi}\ff\nn, \ee where the definition for the parameter
$\sig$ is motivated by the expression for the scalar spectral index
$n_s$ \cite{kinney} \be
n_s-1\simeq\sig-(5-3C)\eps^2-{1\over4}(3-5C)\sig\eps+
{1\over2}(3-C)\xi,\label{ns}\ee with
$C=4(\ln2+\gamma)-5\simeq0.0814514$.  Thus, to lowest order,
$\sig$ describes the departure from scale-invariance.

The flow variables satisfy the following equations (see {\it e.g.}
\cite{kinney} and references therein) \be {d\eps\over dN} &=&
\eps(\sig+2\eps)\ff,\nn\\ {d\sig\over dN} &=&
-\eps(5\sig+12\eps)+2\xi\ff,\label {floweq}\\ {d\over
dN}\ff^\ell\lambda_H &=& \Big[{\ell-1 \over 2}\sig
+(\ell-2)\eps\Big]\ff^\ell\lambda_H+ \ff^{\ell+1}\lambda_H\ff,
\ff(\ell\geq2)\nn \ee where $N$ is the number of e-foldings before
inflation ends. These flow equations have been used widely in the
literature to study inflationary dynamics \cite{me1,liddle,wmap1}. For a
particular choice of initial conditions for the flow parameters,
equations \re{floweq} can be integrated until the end of inflation
without relying on any slow-roll assumption. In this way one can
explore the full dynamics of a large number of simple inflationary
models in a relatively model independent way. Nevertheless, in section 4.2, we shall explore another approach to inflation model building based on $\eps(N)$.

Having generated a model of inflation, the spectrum of curvature
perturbations can be calculated  using analytic formulae 
such as those derived in \cite{stewartlyth,gong,martin,habib+}. In this paper, we shall focus the accuracy of the Stewart-Lyth formula \cite{stewartlyth}, which has been widely used to calculate inflationary perturbations. This approach is not necessarily self-consistent if the conditions of slow roll are violated during inflation (which is, of course, allowed by equations \re{floweq}). An alternative approach is to calculate the perturbation spectra numerically using, for example, the Mukhanov formalism \cite{mukhanov, sasaki}. We briefly compare these two approaches.

\subsection{The Stewart-Lyth approximation}

If the inflaton field $\phi$ rolls so slowly that its kinetic energy
is continually dominated by its potential energy, then the power
spectrum of scalar curvature perturbations (we consider only
scalar perturbations in this paper) can be approximated by the standard first-order slow-roll
approximation: 
\be \mc{P}^{1/2}_{\mc
R}(k)\simeq{2\over\M^2}{H^2\over|H'|}\Bigg|_{k=aH}, \label{anon}
\ee 
where primes denote differentiation with respect to  $\phi$. The next
order correction  is given by Stewart and  Lyth \cite{stewartlyth} \be
\mc{P}^{1/2}_\mc{R}(k)\simeq
\left(1-\eps+{C-3\over8}\sig\right){2\over\M^2}{H^2\over|H'|}\Bigg|_{k=aH},\label{spectrumR}\ee
where  $C$ is the constant introduced in equation \re{ns}. Equations \re{anon}
and \re{spectrumR}  are
evaluated at  the instant when  each Fourier mode $k$ crosses  the Hubble
radius.  

For numerical purpose, we express the Stewart-Lyth power spectrum
purely in terms of the flow parameters: \be\!\!\!
\!\!\!\!\!\!\!\!\!\!\!\!\!\!\!\!\!\!\mc{P}_\mc{R}(k)\simeq\mc{P}_\mc{R}(k_0){
\left[1-\eps+{C-3\over8}\sig\right]^2\eps^{-1}\bigg|_{k=aH}\over\left[1-\eps+{C-3\over8}\sig\right]^2\eps^{-1}\bigg|_{k_0=aH}}\exp\left(-2\int^{\ln|k/k_0|}_{0}\eps(60-N')dN'\right),\label{bigPR}\ee
where $k_0=0.002$ \mpc is the pivot scale at which the spectrum is
normalised.  This normalisation is estimated as \be
\mathcal{P_R}(k_0=0.002\mbox{ Mpc}^{-1})\simeq2.36\times10^{-9}\ff.\ee
from the 3-year WMAP results combined with several other datasets
\cite{wmap3}.

\subsection{The Mukhanov formalism}

The Stewart-Lyth formula assumes that the background parameters $\eps,
\sig$ are approximately constant as each mode crosses the Hubble
radius. However, as soon as the inflaton potential becomes steep,
$\eps$ and $\sig$ will change rapidly and the formula becomes
unreliable. To calculate the curvature spectrum accurately in such 
cases, we compute a numerical solution for the Mukhanov
variable $u$ \cite{mukhanov,sasaki}
\be u = -z\mc{R} \ff, \qquad\ff z= {a\over H}{d\phi\over dt}\ff,\label{ms}\ee
where $\mc{R}$ is the curvature perturbation.  In Fourier space, $u_k$ evolves according to a Klein-Gordon equation with a time-varying effective mass:
\be {d^2 u_k\over d\tau^2}+\bkt{k^2-{1\over z}{d^2 z\over d\tau^2}}u_k=0,\label{KG}\ee where $\tau$ is the conformal time $d\tau = dt /a$.
At early times $\tau_i$, when the mode $k$ is much smaller than the Hubble
 radius, $u_k$ is usually chosen to represent the Bunch-Davies 
vacuum state\footnote{Some authors have suggested that trans-Planckian
physics can be modelled by choosing a different vacuum state \cite{easther,danielsson}}
with
\be u_k(\tau_i)={1\over\sqrt{2k}}e^{-ik\tau_i}.\ee

During inflation, $u_k$ evolves in time until the physical scale of
each mode is far outside the Hubble radius, at which point $|\mc{R}|$
approaches the asymptotic value and freezes.  The power spectrum can
then be evaluated as: \be \mc{P}_\mc{R}(k)={k^3\over
2\pi^2}\lt|{u_k\over z}\rt|_{k\ll aH}^2.\ee

To numerically integrate equation \re{KG}, we rewrite it in terms of
the number of e-foldings $N$ rather than conformal time \be {d^2 u_k
\over dN^2}-(1-\eps){d u_k\over dN}+\lt[\bkt{k\over
aH}^2-f(\eps,\sig,\xi)\rt]u_k=0,\label{niceode}\ee since we use the
inflationary flow equations to calculate the evolution of the
unperturbed background. The factor $f$ in equation \re{niceode} is
given by \be
f(\eps,\sig,\xi)=2-4\eps-{3\over2}\sig-2\eps^2+{\sig^2\over4}+\xi,\ee
and is exact despite its dependence on the flow variables.  The
initial conditions assigned to $u_k$ are: \be
\mbox{Re}(u_k(\tau_i))={1\over\sqrt{2k}}\ff, \qquad
&\mbox{Re}&\bkt{{du_k\over dN}(\tau_i)}= 0\ff,\nn\\
\mbox{Im}\bkt{u_k(\tau_i)}=0\ff, \qquad &\mbox{Im}&\bkt{{du_k\over
dN}(\tau_i)}=-{1\over aH}\sqrt{k\over2}\ff,\ee where $aH/k$ is a small
fixed ratio, which we have set to be $10^{-3}$. We have checked that
the final power spectrum shows negligible change if this ratio is
reduced further.

 For each inflation model, we continue evolving each mode until the
end of inflation, so that \be \mc{P}_\mc{R}(k)=
\mc{P}_\mc{R}(k_0)\lt({k\over k_0}\rt)^3\lt|{u_k\over
u_{k_0}}\rt|^2\sub{end}\label{pmukh}.\ee 
As long as inflation satisfies slow-roll conditions ($\epsilon$, $\sigma
\ll 1$), the numerical solution \re{pmukh} will agree accurately with the
Stewart-Lyth approximation \re{spectrumR} for all modes that are
well outside the horizon at the end of inflation. However, there
can be large differences if the slow conditions are violated temporarily
during inflation. These types of models have been analysed by 
Leach \etal \cite{leachliddle,leach+}. In particular, it was shown in reference \cite{leach+} that the Stewart-Lyth formula will fail if the quantity $z$ in \re{ms}
becomes smaller than its value at the time a mode crosses the Hubble
radius. Some specific examples of this behaviour are described in the
next section.


\section{Limitations of slow-roll approximation}

As in previous papers \cite{kinney,me1,liddle} we use the inflationary flow equations \re{floweq}
to evolve a large number of inflation models with initial flow parameters
selected at random from the uniform distributions
\be \eps_0&\in&[0,0.8]\ff,\nn \\\sig_0&\in&[-0.5,0.5]\ff,\nn \\\C_0&\in&[-0.05,0.05]\ff, \label{window}\\ \LL|_0&\in&[-0.025\times5^{-\ell+3},0.025\times5^{-\ell+3}]\ff, \ff(3\leq\ell\leq10)\nn \\^{11}\lambda_H|_0&=&0\ff.\nn\ee
The distributions \re{window} should not be interpreted as a physical
probability distribution for inflationary models, but merely as a 
device for generating models with a wide range of evolutionary histories. Despite the fact that the truncation is made at $\ell=11$, the types of models that we generated were sufficiently diverse for us to understand when and why the Stewart-Lyth approximation breaks down in general.  Keeping the hierarchy to higher orders can produce more complicated inflationary models, but this will not affect our qualitative conclusions.

We have selected three models for which the Stewart-Lyth approximation
differs to varying degrees from integrating equation \re{KG}.  First,
we consider small scale perturbations where the Stewart-Lyth formula
is inaccurate for modes for which $|\mc{R}|$ does not have time to
reach its asymptotic value well before inflation ends. Next, 
we describe an  inflationary scenario where $\eps$ changes rapidly, arising
from a `bumpy' potential. Finally, we show an example for which
the potential has a steep feature, inducing fast roll and 
interrupting inflation temporarily.  Models based on specific
potentials $V(\phi)$ displaying similar behavior to each of these
three models can be found in 
\cite{leachliddle, leach+, adams}. Throughout, we assume  that there were $60$ e-folds of inflation between the time CMB scale perturbations were generated and the end of inflation.

\subsection{Small-scale perturbations}\label{s1} 

\begin{figure}
\begin{tabular}{cc}
\hskip-0.4in{\includegraphics[width= 3.6in]{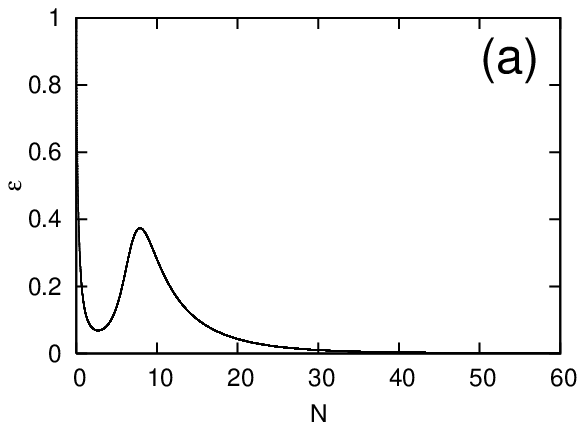}} &
\hskip -0.4in {\includegraphics[width=3.6in]{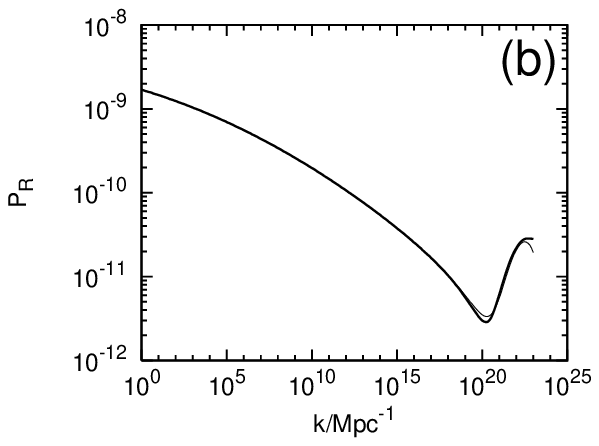}}\\
\end{tabular}
\caption{Panel (a) shows the parameter $\eps$ as a function of $N$, the number of e-folds of inflation (with $N=0$ at the end of inflation). Rise in $\eps$ towards unity at the end of inflation gives rise to the small scale feature of the power spectrum $\pr(k)$, shown in panel (b). The spectrum in bold line uses the Mukhanov variable \re{pmukh} while the thin line shows the Stewart-Lyth approximation \re{bigPR}.}
\label{small}
\end{figure}

Figure \ref{small}a shows  $\eps(N)$ for a typical
inflationary scenario that lasts for 60 e-folds 
generated from the distribution
\re{window}. Within the last few e-folds of inflation, $\eps$ rises
rapidly to unity. Consequently, the damping term in equation
\re{niceode} is suddenly reduced, and $u_k$ evolves more like a
frictionless oscillator. This results in an upturn in the power
spectrum $\pr$ (figure \ref{small}b), peaking at the smallest scales
that exit the Hubble radius.

\begin{figure}
\begin{tabular}{cc}
\hskip-0.4in{\includegraphics[width= 3.6in]{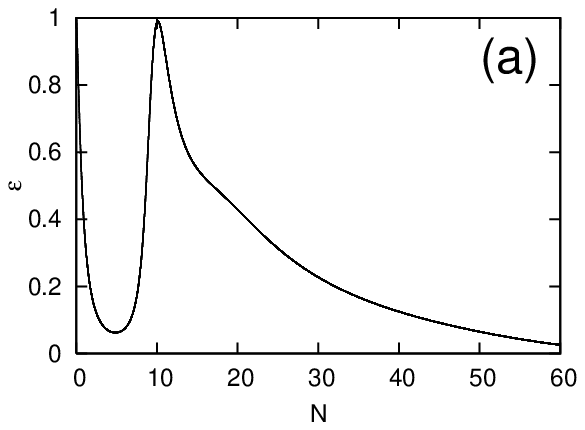}} &
\hskip -0.4in {\includegraphics[width=3.6in]{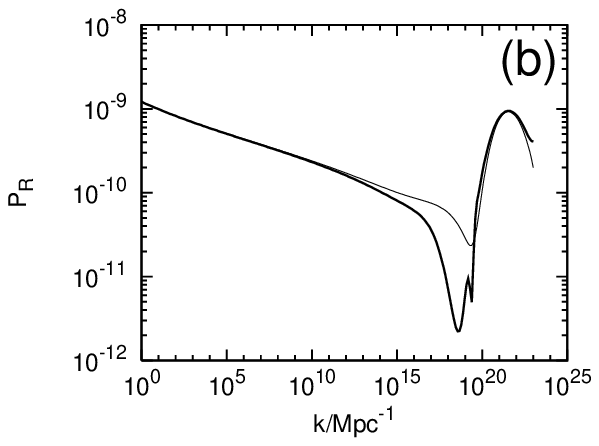}}\\
\end{tabular}
\caption{Panel (a) shows $\eps(N)$ for a model in which $\eps(N)$ peaks prominently during inflation as a consequence of a bumpy potential with a pronounced bump or drop. The origin of the dip in the Mukhanov spectrum (panel (b), bold line) and its magnitude in comparison with that of the Stewart-Lyth spectrum are discussed in the text. }
\label{bumpy}
\end{figure}

\begin{figure}
\begin{tabular}{cc}
\hskip-0.4in{\includegraphics[width= 3.6in]{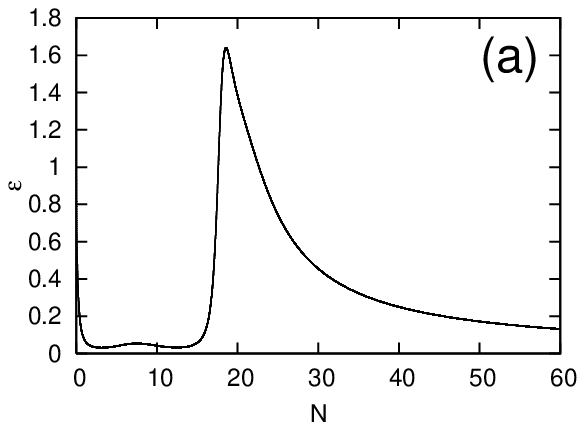}} &
\hskip -0.4in {\includegraphics[width=3.6in]{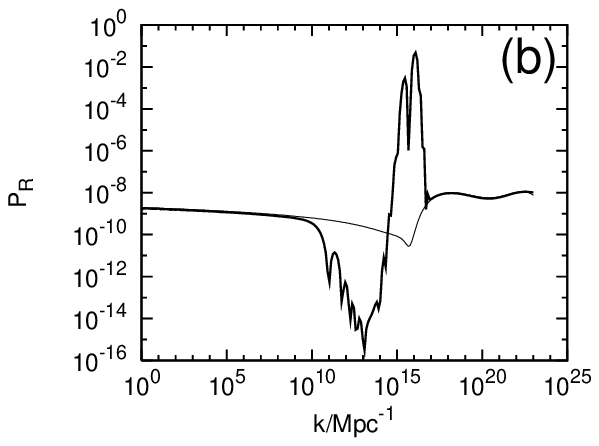}}\\
\end{tabular}
\caption{Panel (a) shows  $\eps(N)$ for a model which temporarily fast-rolls (when $\eps>1$), giving rise to a typical `ringing' in the power spectrum shown in panel (b).   The origin of the large discrepancy between the Mukhanov (bold) and the Stewart-Lyth spectra in panel (b) is discussed in the text. }
\label{fast}
\end{figure}

The numerical solution of the Mukhanov equation \re{KG} (bold line)
and the Stewart-Lyth formula are almost indistinguishable for
wavenumbers $k \lesssim 10^{18}$ \mpc.  Deviations in the solutions can
be seen at larger wavenumbers, but they are relatively small.  This
behaviour is quite typical for models in which inflation ends when
$\epsilon$ exceeds unity. For most purposes the Stewart-Lyth formula
will provide an excellent approximation to the power spectrum even on small
scales. Large deviations from the Stewart-Lyth formula require more 
unusual inflationary models such as the two examples described next.

\subsection{`Bumpy' potential}\label{s2}

Figure \ref{bumpy}a illustrates a scenario in which $\eps$ peaks
strongly during inflation as result of a potential with a pronounced
bump or drop. Here the Stewart-Lyth approximation and the exact
integration agree at large scales but differ considerably as the
background evolves rapidly leading to the prominent dip in $\pr$ as
seen in figure \ref{bumpy}b. We can understand this feature as
follows.  As the background value of $\eps$ rises around a potential
bump during inflation, the dynamics of the flow variables require that
$\sig$ simultaneously plummets to large negative values. Thus, according to the slow-roll formula for the spectral
index \re{ns}, the gradient of the power spectrum also decreases to
large negative values. This explains the downturn in the Stewart-Lyth
spectrum.  However, superhorizon evolution, which is neglected by the
Stewart-Lyth approximation, becomes important when the background
changes rapidly. In this case, as $\sig$ becomes large and negative,
one finds that the Mukhanov variable evolves as $u_k\sim e^{-|A|N}$
where $A=O(\sig)$. This results in an additional decrease in $\pr$
around the bump. Subsequently, towards the true end of inflation, the
tail of the spectrum resembles that in figure \ref{small}b as
expected.

\subsection{Fast roll}\label{s3} 

Consider a scenario in which $\eps(N)$ peaks so strongly during inflation that $\eps$ exceeds unity temporarily, hence breaking up inflation into two or more stages (see \eg \cite{burgess+}). Such a `fast-roll' model is shown in figure \ref{fast}. 

 

The power spectrum resulting from fast roll (figure \ref{fast}b) shows
a large difference between the Mukhanov and the Stewart-Lyth
spectra. In addition to the enhanced dip as explained earlier, we draw
attention to two other features: i) the subsequent sharp rise of the
Mukhanov spectrum several orders of magnitude above the Stewart-Lyth
approximation, and ii) the rapid oscillating features across
$10^{11}\leq k/\mbox{Mpc}^{-1}\leq10^{16}$. 

These features can be understood as follows.  Equation \re{niceode}
clearly shows what happens to the Mukhanov variable $u_k$ during
fast roll. The $(1-\eps)$ friction term of the harmonic oscillator
becomes negative during fast roll, hence boosting $|u_k|$
momentarily. Physically, fast-rolling leads to an entropy perturbation
which sources growth of curvature perturbations on superhorizon scales
\cite{leachliddle}.  The finer-scale `ringing' induced during fast roll
 have previously been seen in  \cite{starobinskii,adams} in their investigations of a potential with a step.

\section{Application to primordial black holes}

Many authors have speculated on the possible existence of primordial
black holes generated during the early Universe
\cite{hawking,zeldovich,carrhawking,khlopov}. Such objects, if they are more
massive than $10^{15}$ g could contribute to the dark matter density
at the present time \cite{afshordi,blais2} and may be the source of
high energy cosmic rays \cite{carrmacgibbon,barrau+} (see \cite{carr2} for a
recent review).  However, it has proved difficult to think of
compelling scenarios in which an interesting density of primordial
black holes could form. Formation of primordial black holes in
inflationary models have been discussed by a number of authors
(see \cite{carrlidsey,yokoyama} and especially \cite{leach2+}, which is the closest in spirit to this work).  We revisit this problem in this section and show that
significant primordial black hole formation during inflation could
have occured only if the inflationary dynamics were highly
unusual and finely tuned. For such unusual inflationary  models, an accurate
computation of the primordial black hole abundance requires 
numerical integration of the inflationary perturbations as described in the
previous section.

\subsection{Basic quantities}

Here, we consider black holes that form when inflation ends and
primordial density fluctuations begin to re-enter the Hubble radius.
In this section we review the basic quantities relevant to the
formation of primordial black holes and compute their relic abundance
using the Sheth-Tormen mass function \cite{sheth}.

Fluctuations in the energy density smoothed over a spherical region of scale 
 $R$ are related to the energy density field $\delta(\mb{x})$ by the
convolution 
\be \dt(R,\mb{x})= \int
W(|\mb{x'}-\mb{x}|/R)\dt(\mb{x'})d^3x',\ee where 
$W$ denotes a `window function' of scale $R$ .  In Fourier space,
the variance of $ \dt(R,\mb{x})$ is  given by
\be \sigma^2(R)&=&\int_0^\infty \mc{P}_\delta(k) \hat W^2(kR){dk\over k},\label{variance}\ee
where $\hat W(kR)$ is the Fourier transform of the window function $W$. 
Here we will assume that $W$ is a top-hat function of radius $R$ (see {\it e.g.}
\cite{llbook}, \S4.3.3).

The power spectrum of the energy density perturbations
$\mc{P}_\delta(k)={k^3/2\pi^2}\langle|\dt(k)|^2\rangle$  
is related to the spectrum of curvature perturbations $\pr(k)$  by \cite{llbook}:

\be \mc{P}_\dt(k)\simeq{16\over81}\bkt{k\over aH}^4\mc{P_R}(k).\ee


A region with density contrast $\dt$ forms a black hole if $\dt$
exceeds a critical value $\dc$.  Previous authors have shown that
$\dc$ should be roughly equal to the equation of state during radiation
era $(\dc\simeq1/3)$ \cite{carrhawking,carr,nadezhin}, and we shall
take $\dc=1/3$ throughout this paper.\footnote{but see
\cite{niemeyer,musco, hawke} for differerent opinions on the value of
$\dc$ inferred from simulations of gravitational collapse.} In
addition to the lower limit there is an upper bound for $\dt$ of
approximately unity, since for such high overdensities large positive
curvatures will cause patches to close on themselves, creating disconnected
`baby' universes \cite{carrhawking}.

Next, to obtain the mass spectrum of primordial black holes, consider
the mass $M$ associated with a region of comoving size $R$ and average
energy density $\rho$.  Early work by Carr \cite{carrreview} suggested 
that primordial black holes form with mass comparable to the 
`horizon mass', \ie \be M\simeq {4\pi\over3}
w^{3/2}\rho\bkt{H^{-1}}^{3}\label{mh}.\ee Recent numerical investigations
\cite{niemeyer}, however , suggest that $M$ may be a small fraction
of the horizon mass, and may be bounded from below
\cite{hawke,chisholm}. Nevertheless, these results have been criticised by 
 some authors \cite{musco}, and we shall still use equation \re{mh} in this paper.
It then follows that $M$ and $R$ are related by
\cite{avelino}

\be{R\over 1 \mbox{ Mpc}}\simeq2.2\times 10^{-24}\lt({M\over 1\mbox{ g}}\rt)^{1/2}\lt({g_*\over2}\rt)^{1/12},\label{massvariance}\ee where the effective relativistic degree of freedom $g_*$ is of order 100 in the very early universe when the temperature exceeds $\sim300$ GeV.  By combining Equations \re{variance}-\re{massvariance}, the variance $\sig(R)$ can then be converted to the mass variance $\sig(M)$.

The fraction of the universe, $\opbh$, in primordial black holes of
mass greater than $M$ is \be
\opbh\equiv{\rho\sub{pbh}\over\rho\sub{total}}
\simeq\int_{\dc}^{1}P(\dt(M))d\dt(M),\ee where $P(\delta)$ is the
density fluctuation probability distribution smoothed on scale $M$ and
is such a sharply falling function that the upper limit can be
extended to infinity without loss of accuracy.  Previous work
\cite{blais,bringmann,sendouda} has used the Press-Schechter
formalism, in which $P(\delta)$ is a Gaussian function, yielding \be
\Omega\sub{pbh}(>M)=\erfc\bkt{\dt\over \sqrt{2}\sig(M)}\label{ps}\ee
Here we use the Sheth-Tormen mass function \cite{sheth} which is based
on an ellipsoidal model of gravitational collapse and provides a
better fit than the Press-Schechter function to the abundances of
non-linear structures in numerical simulations \cite{shethmo}. The
primordial black hole abundance now becomes:

\be
\Omega\sub{pbh}(>M)=\mc{A}\lt[\erfc\lt(\sqrt{a\over2}{\dc\over\sig(M)}\rt)+{1\over
2^p\sqrt{\pi}}\Gamma\lt({1\over2}-p,{a\dc^2\over2\sig^2(M)}\rt)\rt],\label{ost}\ee
where $\mc{A}\simeq0.3222$, $p\simeq0.3$, $a=0.707$ and $\Gamma(x,y)$
is the incomplete gamma function.

\subsection{Results}

For inflationary models that support primordial
black hole production, we calculated the black hole abundance using
the Sheth-Tormen formula\footnote{In calculating $\opbh$, the Sheth-Tormen and the Press-Schecter formalisms broadly agree. However, this depends sensitively on the shape of the inflaton potential. We found classes of potentials where the two prescriptions differ by a few orders of magnitude. Nevertheless, the resulting bound on $n_s$ \re{nsbound} remains valid regardless of which prescription is used.} \re{ost} with $M=10^{15}$g, corresponding to
the lightest black holes that may survive till present day without
evaporating completely. For models that are radiation
dominated at the end of inflation (\ie do not experience an
extended matter dominated phase), the
observational constraints on the primordial black hole abundance can
be described roughly by 
\be
\Omega\sub{pbh}(M>10^{15} \;{\rm g}) \lesssim 10^{-20}. \label{obs1}
\ee
More accurately, the constraint on $\opbh$ is a function of the mass of
the \pbhs and on the physics of the (p)reheating phase \cite{green,greenmalik}. In this
paper it will be adequate to use the simple criterion \re{obs1} to give a
feel for the types of inflation models that might produce black holes.

\subsubsection{Models with blue spectra}

A viable model of inflation requires at least $60$ e-folds of
inflation \cite{liddleleach}.  Using the inflationary flow equations, we find that the vast majority of inflationary
flow models that experience such a large number of e-folds are models
in which the inflaton is trapped in a local minimum of the
potential \cite{kinney,me1}. For such models, inflation never ends unless some
additional physical mechanism is included. Examples of such mechanisms
include an auxiliary field, as in hybrid inflation \cite{lythriotto}
or open strings becoming tachyonic in brane inflation models
\cite{sarangitye}. In these models, the resulting spectrum of
curvature perturbation is very well approximated by a power-law
relation: \be \pr(k) = \pr(k_0)\bkt{k\over
k_0}^{n_s-1},\label{powerlaw}\ee with spectral index $n_s$ given by equation \re{ns}. As the inflaton sits in the local minimum, $\eps$ remains very small while $\sig$ is positive and may be large. Thus, the power spectrum for this type of models is `blue' ($n_s>1)$.  The running of $n_s$ is given by

\be{d n_s\over d\ln k}\simeq -{1\over {1-\eps}}\bkt{2\xi-28\eps^2-5\eps\sig-\bkt{3-5C\over 2}\eps\xi + \bkt{3-C\over4}\xi\sig} .\ee Because the leading order comes from the small $\xi$ term, the running in these blue models is  negligible. Therefore, the shape of the curvature power spectrum in these models can be characterized accurately by a constant spectral index $n_s$. 

\begin{figure}
\begin{center}
\includegraphics[angle=-90,width=4in]{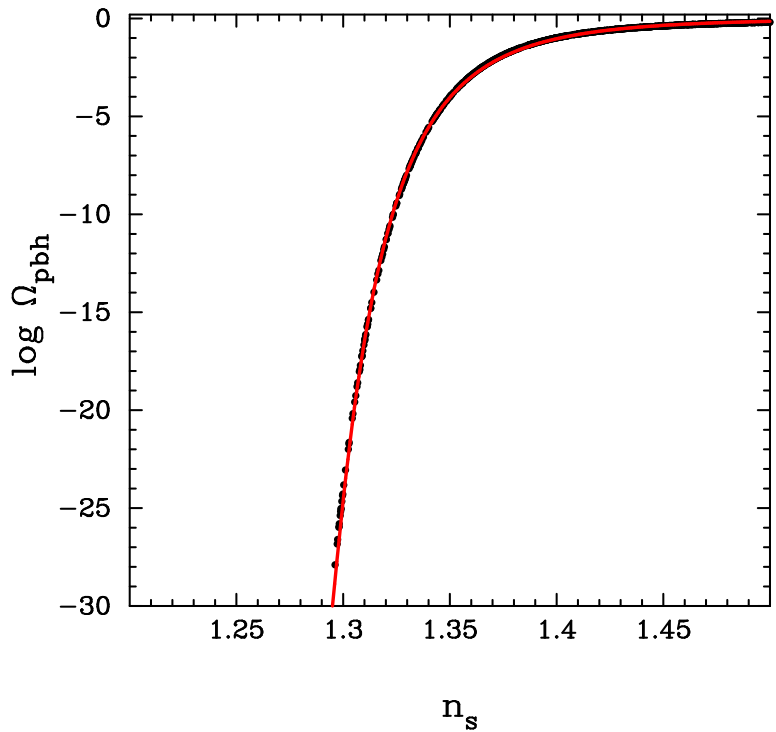}
\end{center}
\caption{Plot of the primordial black hole abundance
$\opbh(M>10^{15}$g) against the spectral index $n_s$ for $10^6$ models
where inflation ends suddenly. There is virtually no dispersion
around the curve given
approximately by equation \re{fit}. The exponential sensitivity of $\opbh$ on $n_s$ at the steep part of the curve highlights the level of fine-tuning needed to produce an interesting density of \pbh.}
\label{bluefig}
\end{figure}

Consequently, for models with blue spectra, $n_s$ directly determines the abundance of primordial
black holes $\opbh$, as shown in Figure \ref{bluefig}. The points show
models trapped in a local minimum of the potential for which inflation
was arbitrarily terminated after 200 e-folds.  The curve shows the
fitting function \be \log_{10}\opbh=-\alpha\bkt{1+\bkt{\beta
(n_s-1.295)}^{p}}^{-1/q},\label{fit}\ee where $\alpha=30$, $\beta=10.2891$,
$p=1.05307$ and $q=0.218219$. Thus, the constraint
$\opbh\lesssim10^{-20}$ discussed above implies an upper bound in $n_s$ \be n_s
\lesssim 1.3, \label{nsbound}\ee in excellent agreement with the analysis of
\cite{green}, which used constraints on the amplitude of the scalar
fluctuations derived from the four-year COBE data \cite{cobe}.  However, the observational constraints from the
3-year WMAP and SDSS \cite{wmap3,seljak,kinney+,peiris} can now convincingly exclude
a high density of primordial black holes from this type of model (for
example, reference \cite{seljak} quotes $n_s = 0.964$ with a $1 \sigma$ error of
only $\pm 0.012$).

\subsubsection{Designing potentials with $\eps$(N)}

The abundance of \pbhs is closely linked to the power spectrum $\pr(k)$ on small scales. 
If the power spectrum has a high amplitude at small scales, as in the case of blue spectra described above, a significant density of massive  \pbhs may form. However, using the inflationary flow approach, models that support 60 e-folds of inflation and end with $\eps=1$, are in general, tilted `red' over a large range of scales and therefore  cannot produce primordial black holes. Even the small upturns seen in figures \ref{small}b and \ref{bumpy}b occur at such low amplitudes that $\opbh$ remains negligible.

However, as discussed in \S3.3, for models which experience a period of fast roll, it is possible to generate a large jump in $\pr$ at small scales, and this can lead to primordial black hole production. For such models, it is important to use a numerical computation to calculate the power spectrum rather than the Stewart-Lyth approximation. For example, for the model shown in figure \ref{fast}, the Stewart-Lyth approximation  predicts a negliglble density of \pbhs whereas  integration of the Mukhanov equation gives $\opbh(M>10^{15}g)\simeq{10^{-5}}$.  One can therefore ask if it is possible to construct fast-roll models that can produce an interesting \pbh abundance, while satisfying the current observational constraints on the shape of the power spectrum at large scales. 

We can construct such models easily based on the Hamilton-Jacobi approach by specifying the dynamics of inflation  in terms of  $\eps(N)$ \footnote{Bond J.R., private communication}. We first choose a set of values of $\eps(N)$ with the boundary conditions:

\be \eps(0)=1,\qquad \lim_{N\rightarrow\infty}\eps=0_+ .\ee 
  
The observational constraints on $r$, $n_s$ and its running can be satisfied by adjusting the values of  $\eps$, $d\eps/dN$ and $d^2\eps/dN^2$ at $N=60$. The form of $\eps(N)$ can then be adjusted `by hand' to generate a brief period of fast roll at small values of $N$ to produce a significant density of primordial black holes.

\begin{figure}
\begin{tabular}{cc}
\hskip-0.4in{\includegraphics[width= 3.6in]{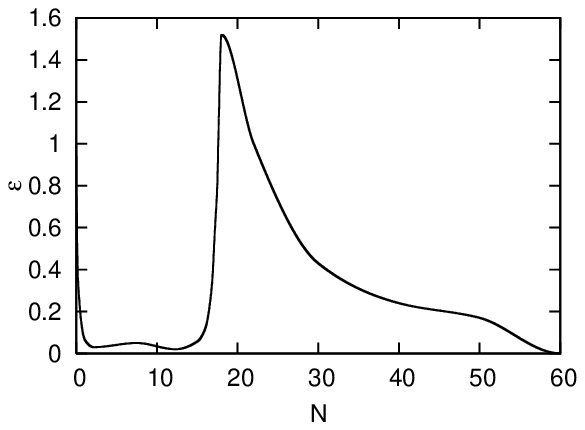}} &
\hskip -0.4in {\includegraphics[width=3.6in]{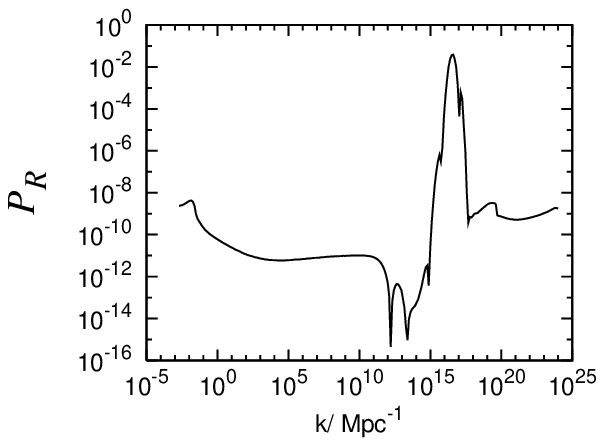}}\\
\end{tabular}
\caption{Left panel shows $\eps(N)$ designed using Hermite interpolating polynomials. Fast-roll regime has been modelled so as to produce \pbhs at an interesting abundance. The reconstructed power spectrum $\pr(k)$ is shown in the right panel. For this model,  $\opbh(M>10^{15}\ff\mbox{g})\simeq1.6\times10^{-20}$. The observables at CMB scales are $r\simeq0.01$, $n_s\simeq0.99$ and $dn_s/d\ln k\simeq10^{-5}$, broadly compatible with constraints from WMAP and other data.}
\label{hermite}
\end{figure}

We then construct $\eps(N)$ by using piecewise cubic Hermite interpolating polynomials. These polynomials are well suited for our purposes since it is easy to apply constraints on the positions of extrema and on the derivatives.  The values of $\eps(N)$ and its first and second derivatives then completely determine the evolution of the Mukhanov variable $u$ via equation \re{niceode}, and thus the power spectrum $\pr$ can be reconstructed. Figure \ref{hermite} shows  $\eps(N)$ for such a fast-roll model and the reconstructed power spectrum. For this model, 
\be\opbh \simeq1.6\times10^{-20}, \ff\ff\ff r\simeq0.01,\ff\ff\ff n_s \simeq 0.99,\ff\ff\ff{dn_s/d\ln k}\simeq 10^{-5},\ee
and so is broadly compatible with observational constraints \cite{wmap3} yet produces a significant abundance of \pbh. It is, of course, possible to generate many variants of this type of model with different values of $n_s$ and $dn_s/ d\ln k$ on CMB scales. But all such models will require a finely tuned period of fast roll towards the end of inflation if they are to produce remnant black holes. 

We end this section with a brief comment on the effect of non-gaussianities on \pbh abundance.  Chen \etal \cite{chen} recently pointed out that step-like features in the inflaton potential can give rise to large non-gaussianities in the primordial perturbations. The effect of non-gaussianities on the abundance of \pbhs was considered in \cite{seery,hidalgo}. These authors found that\footnote{J.-C. Hidalgo, private communication.}:
\be \opbh(>M) \simeq \opbh^{\mbox{\scriptsize{Gaussian}}}+\mc{G}\exp{\bkt{-{\dc^2\over \sig^2(M)}}},\ee where $\mc{G}=\mc{O}(f\sub{NL})$. Thus, even large non-gaussianities anticipated in fast-roll models will be exponentially suppressed, and the overall effect on $\opbh$ is expected to be small.

\section{Conclusions}

There have been many papers discussing the formation of \pbhs from inflation. The simplest such models involve slow-roll inflation and inflationary dynamics that lead to a blue perturbation spectrum. For this type of model, our results confirm earlier work that shows that the scalar spectral index must be less than $n_s\sim1.3$ to avoid overproduction of primordial black holes. However, we also argue that the most natural way to realise such models is to arrange the inflaton to be trapped in a minimum of a potential and for inflation to end abruptly, as in hybrid inflationary models. For such model, the run in the spectral index and the amplitude of the tensor component is negligible. Therefore, observational constraints on the scalar spectral index determined on CMB scales can be extrapolated to the small scales relevant for \pbh production. The strong constraints on the scalar perturbations from WMAP and other data therefore rule out \pbh production from slow-roll inflationary models.

It is, however, possible to generate inflationary models which satisfy observational constraints on CMB scales yet produce a significant density of \pbhs provided there is a period where slow-roll conditions are violated  towards the end of inflation. We have shown in section 3 that for this type of model, it is essential to compute the perturbation power spectra numerically, rather than using slow-roll approximations. Section 4.2 presents a construction method for such a model based on designing $\eps(N)$. While it is possible to produce primordial black holes from inflation, the dynamics of this type of fast-roll model must be very finely-tuned to produce an acceptable density of primordial black holes. In general, even if a model is constructed to have a period of fast roll, the \pbh density will either be excessively high or negligible. Although there has been renewed interest in \pbhs as a possible explanation for the existence of intermediate  mass black holes \cite{mack}, and to account for the rapid growth of supermassive black holes in the centres of galaxies \cite{duechting}, it seems extremely unlikely to us that  primordial black holes formed as a result of inflationary dynamics.

\ack

S.C. is grateful for conversations with Dick Bond, Pascal Vaudrevange, Carlos Hidalgo, Lindsay King and Arantza del Pozo. We also thank Andrew Liddle and Anne Green for  many constructive comments. Our work has been supported by PPARC.

\section*{References}
\bibliographystyle{unsrt}
\bibliography{pbh}

\begin{thebibliography}{10}

\bibitem{albrecht}
A.~Albrecht and P.~J. Steinhardt.
\newblock Cosmology for grand unified theories with radiatively induced
  symmetry breaking.
\newblock {\em Phys. Rev. Lett.}, 48:1220, 1982.

\bibitem{linde}
A.~D. Linde.
\newblock Chaotic inflation.
\newblock {\em Phys. Lett.}, B129:177, 1983.

\bibitem{linde2}
A.~D. Linde.
\newblock Fast-roll inflation.
\newblock {\em JHEP}, 11:052, 2001.

\bibitem{adams}
J.~A. Adams, B.~Cresswell, and R.~Easther.
\newblock Inflationary perturbations from a potential with a step.
\newblock {\em Phys. Rev.}, D64:123514, 2001.

\bibitem{dimopoulos}
K.~Dimopoulos and D.~H. Lyth.
\newblock Models of inflation liberated by the curvaton hypothesis.
\newblock {\em Phys. Rev.}, D69:123509, 2004.

\bibitem{starobinskii}
A.~A. {Starobinsky}.
\newblock {Spectrum of adiabatic perturbations in the universe when there are
  singularities in the inflaton potential}.
\newblock {\em JETP. Lett.}, 55:489, 1992.

\bibitem{matsuda}
T.~Matsuda.
\newblock Elliptic inflation: Generating the curvature perturbation without
  slow-roll.
\newblock {\em JCAP}, 0609:003, 2006.

\bibitem{hoffman}
M.~B. Hoffman and M.~S. Turner.
\newblock Kinematic constraints to the key inflationary observables.
\newblock {\em Phys. Rev.}, D64:023506, 2001.

\bibitem{me2}
S.~Chongchitnan and G.~Efstathiou.
\newblock Prospects for direct detection of primordial gravitational waves.
\newblock {\em Phys. Rev.}, D73:083511, 2006.

\bibitem{kinney}
W.~H. Kinney.
\newblock Inflation: Flow, fixed points and observables to arbitrary order in
  slow roll.
\newblock {\em Phys. Rev.}, D66:083508, 2002.

\bibitem{me1}
S.~Chongchitnan and G.~Efstathiou.
\newblock Dynamics of the inflationary flow equations.
\newblock {\em Phys. Rev.}, D72:083520, 2005.

\bibitem{liddle}
A.~R. Liddle.
\newblock On the inflationary flow equations.
\newblock {\em Phys. Rev.}, D68:103504, 2003.

\bibitem{wmap1}
H.~V. Peiris et~al.
\newblock First year wilkinson microwave anisotropy probe (wmap) observations:
  Implications for inflation.
\newblock {\em Astrophys. J. Suppl.}, 148:213, 2003.

\bibitem{stewartlyth}
E.~D. Stewart and D.~H. Lyth.
\newblock A more accurate analytic calculation of the spectrum of cosmological
  perturbations produced during inflation.
\newblock {\em Phys. Lett. B}, 302:171, 1993.

\bibitem{gong}
J.-O. Gong and E.~D. Stewart.
\newblock The density perturbation power spectrum to second-order corrections
  in the slow-roll expansion.
\newblock {\em Phys. Lett.}, B510:1, 2001.

\bibitem{martin}
J.~Martin and D.~J. Schwarz.
\newblock Wkb approximation for inflationary cosmological perturbations.
\newblock {\em Phys. Rev.}, D67:083512, 2003.

\bibitem{habib+}
S.~Habib, K.~Heitmann, G.~Jungman, and C.~Molina-Paris.
\newblock The inflationary perturbation spectrum.
\newblock {\em Phys. Rev. Lett.}, 89:281301, 2002.

\bibitem{mukhanov}
V.~F. Mukhanov.
\newblock Quantum theory of gauge-invariant cosmological perturbations.
\newblock {\em Sov. Phys. JETP}, 67:1297, 1988.

\bibitem{sasaki}
M.~Sasaki.
\newblock Large scale quantum fluctuations in the inflationary universe.
\newblock {\em Prog. Theor. Phys.}, 76:1036, 1986.

\bibitem{wmap3}
D.~N. Spergel et~al.
\newblock Wilkinson microwave anisotropy probe (wmap) three year results:
  Implications for cosmology.
\newblock 2006.

\bibitem{easther}
R.~Easther, B.~R. Greene, W.~H. Kinney, and G.~Shiu.
\newblock Inflation as a probe of short distance physics.
\newblock {\em Phys. Rev.}, D64:103502, 2001.

\bibitem{danielsson}
U.~H. Danielsson.
\newblock A note on inflation and transplanckian physics.
\newblock {\em Phys. Rev.}, D66:023511, 2002.

\bibitem{leachliddle}
S.~M. Leach and A.R. Liddle.
\newblock Inflationary perturbations near horizon crossing.
\newblock {\em Phys. Rev.}, D63:043508, 2001.

\bibitem{leach+}
S.~M. Leach, M.~Sasaki, D.~Wands, and A.~R. Liddle.
\newblock Enhancement of superhorizon scale inflationary curvature
  perturbations.
\newblock {\em Phys. Rev.}, D64:023512, 2001.

\bibitem{burgess+}
C.~P. Burgess, R.~Easther, A.~Mazumdar, D.F. Mota, and T.~Multamaki.
\newblock Multiple inflation, cosmic string networks and the string landscape.
\newblock {\em JHEP}, 0505:067, 2005.

\bibitem{hawking}
S.~Hawking.
\newblock Gravitationally collapsed objects of very low mass.
\newblock {\em Mon. Not. Roy. Astron. Soc.}, 152:75, 1971.

\bibitem{zeldovich}
Ya.~B. Zeldovich and I.D. Novikov.
\newblock The hypothesis of cores retarded during expansion and the hot
  cosmological model.
\newblock {\em Sov. Astron. A.J.}, 10:602, 1967.

\bibitem{carrhawking}
B.~J. Carr and S.~W. Hawking.
\newblock Black holes in the early universe.
\newblock {\em Mon. Not. R. Astr. Soc.}, 168:399, 1974.

\bibitem{khlopov}
M.~Y. {Khlopov} and A.~G. {Polnarev}.
\newblock {Primordial black holes as a cosmological test of grand unification}.
\newblock {\em Physics Letters B}, 97:383, December 1980.

\bibitem{afshordi}
N.~Afshordi, P.~McDonald, and D.~N. Spergel.
\newblock Primordial black holes as dark matter: The power spectrum and
  evaporation of early structures.
\newblock {\em Astrophys. J.}, 594:L71, 2003.

\bibitem{blais2}
D.~Blais, C.~Kiefer, and D.~Polarski.
\newblock Can primordial black holes be a significant part of dark matter?
\newblock {\em Phys. Lett.}, B535:11, 2002.

\bibitem{carrmacgibbon}
B.~J. Carr and J.~H. MacGibbon.
\newblock Cosmic rays from primordial black holes and constraints on the early
  universe.
\newblock {\em Phys. Rept.}, 307:141, 1998.

\bibitem{barrau+}
A.~Barrau, G.~Boudoul, and L.~Derome.
\newblock An improved gamma-ray limit on the density of pbhs.
\newblock 2003.

\bibitem{carr2}
B.~J. Carr.
\newblock Primordial black holes: Do they exist and are they useful?
\newblock 2005.

\bibitem{carrlidsey}
B.~J. Carr and James~E. Lidsey.
\newblock Primordial black holes and generalized constraints on chaotic
  inflation.
\newblock {\em Phys. Rev.}, D48:543, 1993.

\bibitem{yokoyama}
J.~{Yokoyama}.
\newblock {Formation of Primordial Black Holes in Inflationary Cosmology}.
\newblock {\em Progress of Theoretical Physics Supplement}, 136:338, 1999.

\bibitem{leach2+}
S.~M. Leach, I.~J. Grivell, and A.~R. Liddle.
\newblock Black hole constraints on the running mass inflation model.
\newblock {\em Phys. Rev.}, D62:043516, 2000.

\bibitem{sheth}
R.~K. Sheth and G.~Tormen.
\newblock Large scale bias and the peak background split.
\newblock {\em Mon. Not. Roy. Astron. Soc.}, 308:119, 1999.

\bibitem{llbook}
A.~R. Liddle and D.~H. Lyth.
\newblock {\em Cosmological inflation and large-scale structure}.
\newblock Cambridge University Press, Cambridge, 2000.

\bibitem{carr}
B.~J. Carr.
\newblock The primordial black hole mass spectrum.
\newblock {\em Ap. J.}, 201:1, 1975.

\bibitem{nadezhin}
D.~K. Nadezhin, I.~D. Novikov, and Polnarev A.G.
\newblock The hydrodynamics of primordial black hole formation.
\newblock {\em Sov. Astron.}, 22:129, 1978.

\bibitem{niemeyer}
J.~C. Niemeyer and K.~Jedamzik.
\newblock Dynamics of primordial black hole formation.
\newblock {\em Phys. Rev.}, D59:124013, 1999.

\bibitem{musco}
I.~Musco, J.~C. Miller, and L.~Rezzolla.
\newblock Computations of primordial black hole formation.
\newblock {\em Class. Quant. Grav.}, 22:1405, 2005.

\bibitem{hawke}
I.~Hawke and J.~M. Stewart.
\newblock The dynamics of primordial black-hole formation.
\newblock {\em Class. Quantum Grav.}, 19:3687, 2002.

\bibitem{carrreview}
B.~J. Carr.
\newblock {\em Observational and theoretical aspects of relativistic
  astrophysics and cosmology}.
\newblock (Sanz, J. L. and Goicoechea, L.J. eds.), World Scientific, Singapore,
  1985.

\bibitem{chisholm}
J.~R. Chisholm.
\newblock Primordial black hole minimum mass.
\newblock {\em Phys. Rev.}, D74:043512, 2006.

\bibitem{avelino}
P.~P. Avelino.
\newblock Primordial black hole abundance in non-gaussian inflation models.
\newblock {\em Phys. Rev. D.}, 72:124004, 2005.

\bibitem{blais}
D.~Blais, T.~Bringmann, C.~Kiefer, and D.~Polarski.
\newblock Accurate results for primordial black holes from spectra with a
  distinguished scale.
\newblock {\em Phys. Rev.}, D67:024024, 2003.

\bibitem{bringmann}
T.~Bringmann, C.~Kiefer, and D.~Polarski.
\newblock Primordial black holes from inflationary models with and without
  broken scale invariance.
\newblock {\em Phys. Rev.}, D65:024008, 2002.

\bibitem{sendouda}
Y.~Sendouda, S.~Nagataki, and K.~Sato.
\newblock Mass spectrum of primordial black holes from inflationary
  perturbation in the randall-sundrum braneworld: A limit on blue spectra.
\newblock {\em JCAP}, 0606:003, 2006.

\bibitem{shethmo}
R.~K. Sheth, H.~J. Mo, and G.~Tormen.
\newblock Ellipsoidal collapse and an improved model for the number and spatial
  distribution of dark matter haloes.
\newblock {\em Mon. Not. Roy. Astron. Soc.}, 323:1, 2001.

\bibitem{green}
A.~M. Green and A.~R. Liddle.
\newblock Constraints on the density perturbation spectrum from primordial
  black holes.
\newblock {\em Phys. Rev.}, D56:6166, 1997.

\bibitem{greenmalik}
A.~M. Green and K.~A. Malik.
\newblock Primordial black hole production due to preheating.
\newblock {\em Phys. Rev.}, D64:021301, 2001.

\bibitem{liddleleach}
A.~R. Liddle and S.~M. Leach.
\newblock How long before the end of inflation were observable perturbations
  produced?
\newblock {\em Phys. Rev.}, D68:103503, 2003.

\bibitem{lythriotto}
D.~H. Lyth and A.~Riotto.
\newblock Particle physics models of inflation and the cosmological density
  perturbation.
\newblock {\em Phys. Rept.}, 314:1, 1999.

\bibitem{sarangitye}
S.~Sarangi and S.-H.H. Tye.
\newblock Cosmic string production towards the end of brane inflation.
\newblock {\em Phys. Lett.}, B536:185, 2002.

\bibitem{cobe}
C.~L. Bennett et~al.
\newblock 4-year cobe dmr cosmic microwave background observations: Maps and
  basic results.
\newblock {\em Astrophys. J.}, 464:L1, 1996.

\bibitem{seljak}
U.~Seljak, A.~Slosar, and P.~McDonald.
\newblock Cosmological parameters from combining the lyman-alpha forest with
  cmb, galaxy clustering and sn constraints.
\newblock 2006.

\bibitem{kinney+}
W.~H. Kinney, E.~W. Kolb, A.~Melchiorri, and A.~Riotto.
\newblock Inflation model constraints from the wilkinson microwave anisotropy
  probe three-year data.
\newblock {\em Phys. Rev.}, D74:023502, 2006.

\bibitem{peiris}
H.~Peiris and R.~Easther.
\newblock Recovering the inflationary potential and primordial power spectrum
  with a slow roll prior: Methodology and application to wmap 3 year data.
\newblock {\em JCAP}, 0607:002, 2006.

\bibitem{chen}
X-g. Chen, R.~Easther, and E.~A. Lim.
\newblock Large non-gaussianities in single field inflation.
\newblock 2006.

\bibitem{seery}
D.~Seery and J.~C. Hidalgo.
\newblock Non-gaussian corrections to the probability distribution of the
  curvature perturbation from inflation.
\newblock {\em JCAP}, 0607:008, 2006.

\bibitem{hidalgo}
J.-C. Hidalgo.
\newblock To be published.

\bibitem{mack}
K.~J. Mack, J~P. Ostriker, and M.~Ricotti.
\newblock Growth of structure seeded by primordial black holes.
\newblock 2006.

\bibitem{duechting}
N.~D{\"{u}}chting.
\newblock Supermassive black holes from primordial black hole seeds.
\newblock {\em Phys. Rev.}, D70:064015, 2004.

\end{thebibliography}

\end{document}